\documentclass[aps,showpacs,floatfix,superscriptaddress]{revtex4}
\usepackage{graphicx}
\usepackage{subfigure}
\usepackage{amsmath}
\usepackage{amssymb}
\newcommand{\ds}{\displaystyle}
\newcommand{\scs}{\scriptscriptstyle}

\def\<{\langle}
\def\>{\rangle}
\begin{document}
\title{Vibrational entropy and the structural organization of proteins}\date{\today}
\date{\today}

\author{Lorenzo Bongini}
\email{Bongini@fi.infn.it}
\affiliation{Dipartimento di Fisica, Universit\`a di Firenze, 
             Via. G. Sansone 1, 50019 Sesto F.no (FI), Italy}
\affiliation{Centro Interdipartimentale per lo Studio delle Dinamiche Complesse (CSDC),
	 Universit\`a di Firenze, Via. G. Sansone 1, 50019 Sesto F.no (FI), Italy}             
\author{Francesco Piazza\footnote{Francesco Piazza and Lorenzo Bongini contributed equally to this work}}
\email{Francesco.Piazza@epfl.ch}
\affiliation{Ecole Polytechnique F\'ed\'erale de Lausanne,
         Laboratoire de Biophysique Statistique, ITP--SB,  
         BSP-722, CH-1015 Lausanne, Switzerland}
\author{Lapo Casetti}         
\affiliation{Dipartimento di Fisica, Universit\`a di Firenze, 
         Via. G. Sansone 1, 50019 Sesto F.no (FI), Italy}  
\affiliation{Centro Interdipartimentale per lo Studio delle Dinamiche Complesse (CSDC),
	 Universit\`a di Firenze, Via. G. Sansone 1, 50019 Sesto F.no (FI), Italy}
\affiliation{Istituto Nazionale di Fisica Nucleare (INFN), sezione di Firenze,
         Via. G. Sansone 1, 50019 Sesto F.no (FI), Italy}                
\author{Paolo De Los Rios}
\affiliation{Ecole Polytechnique F\'ed\'erale de Lausanne,
         Laboratoire de Biophysique Statistique, ITP--SB,  
         BSP-722, CH-1015 Lausanne, Switzerland}

%
\begin{abstract}

In this paper we analyze the vibrational spectra of a large ensemble of 
non-homologous protein structures by means of a novel tool, that we coin 
the Hierarchical Network Model (HNM). Our coarse-grained scheme accounts for 
the intrinsic heterogeneity of force constants displayed by protein arrangements 
and also incorporates side-chain degrees of freedom. 

Our analysis shows that vibrational entropy per unit residue correlates
with the content of secondary structure. Furthermore, we assess the individual contribution
to vibrational entropy of the novel features of our scheme as compared with the predictions 
of state-of-the-art  network models.
This analysis highlights the importance of properly accounting for the intrinsic 
hierarchy in force strengths typical of the different atomic bonds that build up 
and stabilize protein scaffolds.

Finally, we discuss possible implications of our findings in the context of protein 
aggregation phenomena.

\end{abstract}

%
%
%
\pacs{87.15.-v,87.14.et,87.15.bd}
\keywords{protein dynamics,  coarse-grained network models, vibrational entropy. }
%
%

\maketitle

\bigskip

\noindent\textbf{Author summary} 

\smallskip 

The intricate structure/dynamics/function relations displayed by proteins are at the core
of life itself. Yet, a thorough and general understanding of such interplay still remains a formidable
task, at the frontier of physics and biology. Proteins perform their functions while fluctuating 
at room temperature about their native folds. As a consequence, the entropic contribution to their 
free energy landscapes, i.e. vibrational entropy, constitutes a crucial element to decipher the dynamical bases of 
protein functions. In this study, we examine a whole database of 
highly-non-homologous protein structures in the effort of rationalizing the entropic contributions of
the distinct secondary structure motifs in protein scaffolds.
With the help of a novel coarse-grained model, we measure a significant correlation between secondary 
structure content and vibrational entropy, thus shedding light into the structural roots of protein 
flexibility. Finally, we discuss our findings in the context of the unsolved problem of protein aggregation.

\bigskip 

\section{Introduction}

In its native state a protein fluctuates around a configuration corresponding to
the absolute minimum of its energy landscape. However, realistic energy landscapes
are characterized by many competing minima. Protein folding can therefore be
viewed as the process of selecting the right minimum among many others. Since
proteins perform their function at finite temperatures, the entropy of their native states
concurs in determining their stability even when folding is driven
mainly by an enthalpic bias. When only one basin of the energy landscape is 
significantly visited, as it happens for folded proteins, entropy is mainly contributed by
fluctuations around the minimum --  it is all {\em vibrational} entropy. 

Native states are highly compact structures, characterized by the presence of
secondary structure motifs, such as $\alpha$-helices and  $\beta$-sheets. In turn, the
vibrational properties of a heterogeneous system are typically influenced by the presence of
simmetries and by its degree of modularity. It is therefore interesting to
investigate whether a correlation exists between the presence of $\alpha$ and 
$\beta$ structures and vibrational entropy. Previous work
has shown that both $\alpha$-helices and $\beta$-sheets are characterized by a
larger flexibility than random-coil conformations~\cite{Ma:2000ee}. However, a systematic study quantifying 
the impact of secondary structure content on the vibrational properties of 
native protein conformations is lacking. 

It is the aim of this paper to introduce the simplest coarse-grained model able to quantify
the correlation between vibrational dynamics of proteins and their structural  organization at
the secondary level. A sensible starting point for such analysis is represented by the class
of coarse-grained network models, that map a given protein structure on a network of
point-like aminoacids interacting trough Hookean springs. 

\begin{figure}[t!]
\centering
\subfigure[]{\includegraphics[clip,width=8 truecm]{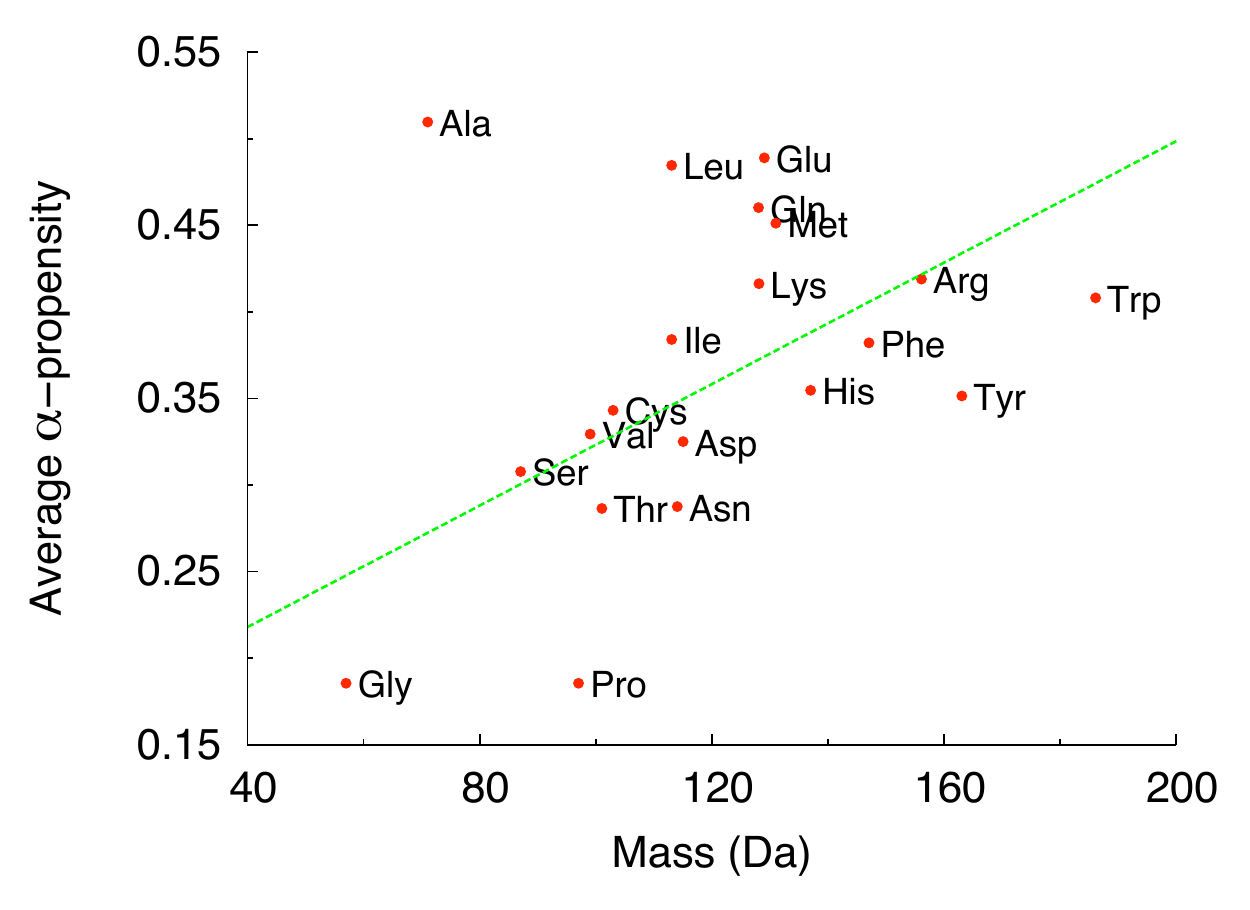}}
\subfigure[]{\includegraphics[clip,width=8 truecm]{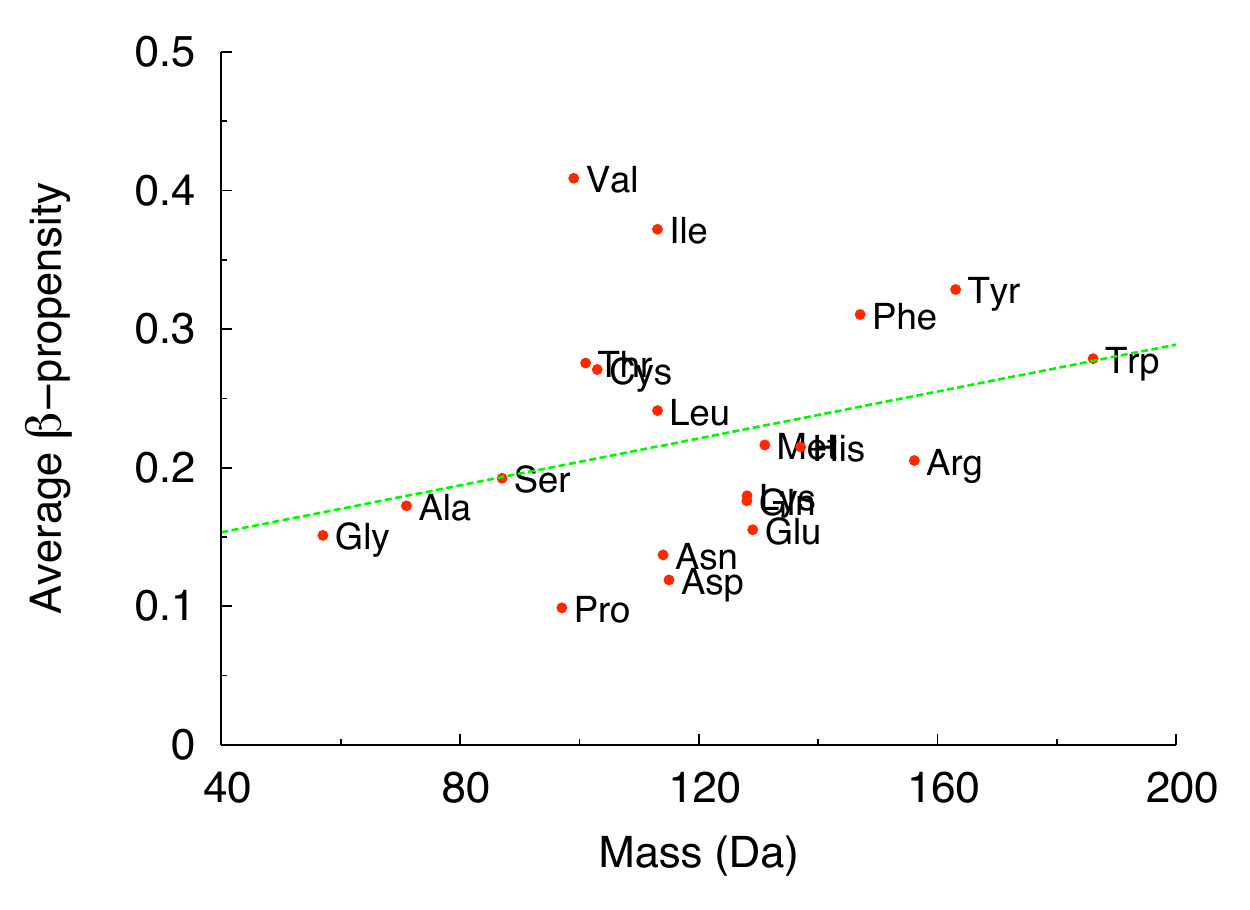}}
\caption{Plot of the statistical $\alpha$ (a) and $\beta$ (b) propensities in the PDBselect database 
(see text). i.e. the normalized frequency of appearance of all residues in a given secondary structure 
motif according to the DSSP protocol~\cite{Wolfgang-Kabsch:1983lr}. The solid lines
are linear fits that correspond to correlation coefficients of 0.44 ($\alpha$) and
0.39 ($\beta$).}
\label{f:masses}
\end{figure}

In most implementations aminoacids are taken to sit at the corresponding C$_\alpha$ 
sites and are assigned the same average mass of 120 Da.
In this framework, a given native structure specifies by construction the
topologies of inter-particles interactions, that is the networks of connectivities
and equilibrium bonds. The patterns of low-frequency collective modes are uniquely
dictated by the topology of the network of physical interactions that characterizes
the native configuration~\cite{Tama:01}. In this sense, network models might be
considered as the simplest tools to describe the low-frequency
regime of protein dynamics. However, due to the coarse-graining element, such
schemes are more questionable in the high-frequency part of the vibrational
spectrum. This makes them unsuitable to compute vibrational entropies, which
include nonlinear contributions from all modes, as it shows from its very
definition

\begin{eqnarray}
\label{anvedilentropiavibbrazzionale!}
TS_\mathrm{vib}  &=&  \sum_i 
                       \frac{\ds h \nu_i}{\ds e^{\beta h \nu_i} -1} 
                     - \beta^{-1} \ln \left( 1-e^{-{\beta h \nu_i}} \right) \nonumber \\
&\simeq& - \beta^{-1}\sum_i \ln \nu_i + \mathrm{const.}
\end{eqnarray}
where the last passage follows in the classical limit $\beta^{-1}=k_{\scs B} T  \gg h \nu_{i} \,$ $\forall \, i$.
As a consequence, one  needs to incorporate further structural and dynamical 
details within the network description besides the sheer topology of connections  
in order to correctly compute vibrational entropies. More precisely, we are interested in 
capturing the structural and dynamical elements that characterize fluctuations at the scale of 
secondary structures. We identify three crucial features that must be taken into account.
\begin{itemize}
\item[$(i)$]   A realistic hierarchy of  force constants, reproducing the 
               different strengths of bonds featured in a protein molecule. 
	       These comprise covalent bonds (such 
               as the peptide bond along the chain), and in particular the interactions 
	       stabilizing secondary 
	       structures. i.e. hydrogen bonds, besides  long-range forces such as 
\item[$(ii)$]   The degrees of freedom  corresponding to side chains, which play a fundamental 
               role in the spatial organization of secondary motifs.	      
	       hydrophobic or screened electrostatic interactions.
\item[$(iii)$] The appropriate aminoacid masses, correctly reproducing the true protein 
               sequence.
               In fact, as shown in Fig.~\ref{f:masses}, residues of different mass show 
	       different $\alpha$ and $\beta$ propensities. In particular light residues tend to be 
	       less represented in secondary structure motifs.	                     
\end{itemize} 

As we will show later, this analysis highlights that $\beta$-rich (and $\alpha$-poor) native conformations tend to 
have a higher vibrational entropy per residue regardless of protein size and shape. 

This is particularly interesting with respect to protein aggregation since protein aggregates are
know to be  rich in $\beta$-structures. In most cases, they share the same $\beta$-spine
architecture  characteristic of amyloid fibrils: an assembly of beta sheets perpendicular to the 
fibril axis. Moreover, aggregation is a rather common phenomenon for all sorts of polypeptide
chains regardless of their sequence~\cite{sequence-determinants}. This suggests that the
phenomenon should be governed by rather general laws, uniquely related to the dynamical properties
of poly-peptide  chains in solution. In particular, in view of the rather slow time-scales
characteristic of  aggregation and in view of our results on the correlation between vibrational
entropy and secondary structure, it is tempting to postulate a mechanism of thermodynamic origin,
that would favor the growth of structures rich in $\beta$-content under quite generic conditions. 

The paper is organized as follows. In the Methods section we describe the database of
native structures and the models used for our analysis. In the Discussion section we discuss the
features of vibrational spectra and the emergence of the observed correlation between vibrational
entropy and  secondary structure. Finally, we comment on the biological relevance of our
results, particularly for what concerns protein aggregation phenomena.

\section{Methods}
\label{Models and methods}
%
\subsection{The database}

It is our aim to conduct the most general analysis of the interplay between vibrational 
entropy and the content of secondary structure in a large database of protein structures.
Of course, we are bound to avoid repeatedly taking into account structures 
corresponding to homologous proteins.  We therefore chose the PDBselect database, 
which was explicitly  built to gather the largest number of protein scaffolds by 
keeping the structural homology  between any two structures lower than 30 \%~\cite{PDBSELECT}.

\begin{figure}[t!]
\centering
\includegraphics[clip,width=12 truecm]{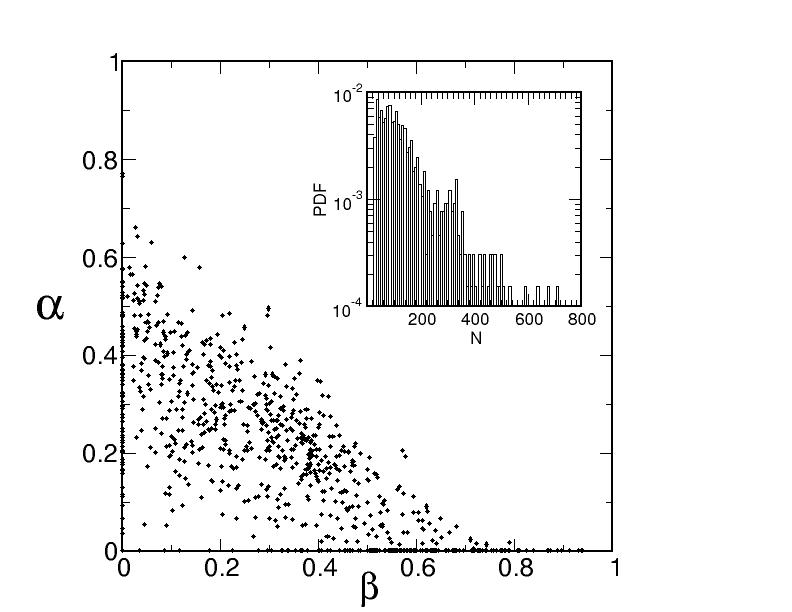}
\caption{$\alpha$ versus $\beta$-content in the PDBselect database. The inset shows the 
histogram of chain lengths in the database (lin-log scale).}
\label{f:alpha-beta}
\end{figure}

In order to illustrate the amount of structural diversity displayed by proteins
from the PDBselect database, we show in Fig.~\ref{f:alpha-beta} both  the $\alpha$-helix and 
$\beta$-sheet content of each structure as computed by means
of the DSSP algorithm introduced by Kabsch and Sanders~\cite{Wolfgang-Kabsch:1983lr}. 
We see  that the two measures roughly
anti-correlate, their sum being Gaussian-distributed around 50 \% 
with a standard deviation of about 15\%. 
As a further piece of information, we report in the inset of Fig.~\ref{f:alpha-beta} 
the statistical distribution  of chain lengths for all sequences contained in the database,
which shows to decay exponentially. Additionally,  we found no appreciable correlation 
between $\alpha$ or $\beta$-content and other structural 
indicators such as chain length or surface accessible area. 
On the contrary, although not surprisingly, the sum of $\alpha$ and 
$\beta$-content correlates positively with the number of hydrogen bonds (HB), 
which is indeed the principal physical interaction stabilizing both kind of motifs.

\subsection{The HNM model}

Our starting point is a widely employed coarse-grained scheme, the Anisotropic Network Model
(ANM). Originally proposed by Tirion~\cite{Tirion:95,Tirion:1996mz} at the all-atom level,
such model has been successively reconsidered by Bahar and co-workers in the C$_\alpha$ 
approximation~\cite{Pemra-Doruker:2000fk}, with results in fair agreement with experiment. In
the spirit of coarse-grained network models, aminoacids are modelled as spherical beads
linked by Hookean springs if they are close enough in the  native structure, as specified
through a pre-assigned cutoff distance.  Both ANM and its scalar analogue, the Gaussian
Network Model (GNM)~\cite{Bahar:1997qy,Eyal:2006uq}, proved very efficient in describing the
low-frequency part of protein spectra, which is dominated by large-scale, collective motions
of entire domains or other structural sub-units, as measured from X-ray crystallography or
electron microscopy~\cite{Brooks:05,Tama:2002gb,Delarue:02,De-Los-Rios:2005qy,Hinsen:05,Hinsen:98},

Let $N$ be the number of residues of a given structure and let $\rho_{ij}$ denote 
the distance between the C$_\alpha$s of the $i$-th and $j$-th  aminoacids in the PDB 
structure~\footnote{In the spirit of elastic network models, the PDB structure is taken
as an approximation of the native structure, i.e. the global minimum of the {\em true} potential 
energy of the system. Hence, no minimization is performed and the network of interactions is
built directly from the PDB coordinates.}.
Then, the network of residue-residue interactions can be constructed by introducing a distance 
cutoff $R_{c}$,  and the corresponding {\em connectivity} matrix 
\begin{equation}
\label{e:connmatr}
\Gamma_{ij} =  
\left\lbrace
\begin{array}{ll}
1 & \rho_{ij} \leq R_{c} \\
0 & \rho_{ij} < R_{c} \\
\end{array}
\right.
\end{equation} 
Accordingly, the total potential energy of the C$_{\alpha}$ ANM is a sum of harmonic potentials,
\begin{equation}
\label{e:ANMpot}
V=\frac{1}{2} \sum_{i<j} k \Gamma_{ij} (r_{ij}-\rho_{ij})^2
\end{equation}
Here $r_{ij}=|\vec{x}_{i}-\vec{x}_{j}|$ measures the instantaneous elongation of 
the $(i,j)$ bond and $k$ is the strength of all springs connecting interacting pairs.

The ANM model is known to describe accurately long-wavelength fluctuations, such as the
concerted motions of subunits or entire domains, whereas some doubts can be cast on its
predictive accuracy concerning aminoacid motions at a scale comparable to the characteristic
dimension of a residue. However, we are interested in computing vibrational entropies, which 
depend on the whole spectrum. Importantly, the contributions of high and low
frequencies to the entropy are impossible to disentangle in an elementary manner, due to its
strong nonlinearity. Hence, in order to increase the spectral reliability of the C$_{\alpha}$
scheme and make it more accurate in the high-frequency domain, we introduce  three additional
features within the standard ANM model.

\begin{enumerate}
\item In the basic ANM protocol, all springs share the same strength, in spite of 
the wide differences among the real forces governing residue-residue interactions in a real protein.
Hence, we introduce a {\em hierarchy} of spring constants, aimed  at reproducing the strength of 
the most important  interactions, that is
covalent, hydrogen-bond and Van der Waals (VW) bonds.
Consequently, we modify the potential energy function~(\ref{e:ANMpot}) as follows
\begin{eqnarray}
\label{e:HNMpot}
V= &\alpha& \sum_{i<j} C_{ij} (r_{ij}-\rho_{ij})^2 +\\
   &\beta&  \sum_{i<j} H_{ij} (r_{ij}-\rho_{ij})^2 +
     \gamma \sum_{i<j} W_{ij} (r_{ij}-\rho_{ij})^2, 
\end{eqnarray}
where $C$, $H$ and $W$ are the connectivity matrices of the three distinct
sub-networks comprising all aminoacids interacting via peptide bonds along the main chain, 
through HBs and VW interactions, respectively. The quantities $\alpha$, $\beta$ and $\gamma$ are the corresponding 
spring constants, whose magnitude is taken by construction to span two orders of magnitude. 
We coin this scheme the Hierarchical Network Model (HNM).

Spring constants customarily used to model covalent bonding are in the range $300-400$ 
kcal/mol/\AA$^2$. Here we chose $\alpha = 300$ kcal/mol/\AA. 
A second order expansion of the Lennard-Jones potentials used to describe HB in different 
all-atom force fields gives values of $\beta$ in the range  $30-90$ kcal/mol/\AA$^2$.
We tried values of $\beta$ in the range $0.1-0.5$ $\alpha$ with no appreciable difference in the distinctive 
features of the vibrational spectra. Finally, taking into account that  
the young modulus of covalent solids is about  $10^3$ times that of a Van der
Waals solid, we assume $\alpha/\gamma=10^3$. 

\item Coarse-graining protein structure at the C$_{\alpha}$ level, one neglects all degrees of freedom associated 
with aminoacid side chains. However, these are known to be subject to different positional constraints in 
secondary structure motifs such as parallel and anti-parallel $\beta$-sheets and $\alpha$-helices. 
Moreover, the entropic contribution of side chains has been recently 
proved fundamental in determining the free energy of protein 
native states~\cite{Daniele-Sciretti:2009ek}
Hence, we also consider a variant of the HNM model where we model  side chains as additional beads, whose equilibrium position
we fix at the side chain center of mass in the native structure. Furthermore, we place a covalent 
bond between each bead-like side chain and  its corresponding $\alpha$ carbon. 
In particular, given their small size, we neglect side chains of Glycines, which will be only represented 
by their $\alpha$ carbon bead. We remark that this  approach is utterly similar to many other 
unified residue models~\cite{Zacharias:2003fk}.

\item Usually, in the framework of protein network models, one assigns to each $\alpha$ carbon bead the average 
aminoacid mass of 120 Da. In the present study, we adopt the following convention:
we split the mass of each  aminoacid into a constant contribution of 56 Da, that represents the 
atoms lying along the peptide bond, and assign a variable mass ranging
from 35 Da (for Alanine) to 150 Da (for Tryptophan) to the bead representing the side chain.

\end{enumerate}

In the following, we will present results for an {\em hybrid} HNM model, whereby the above three features 
shall be switched on and off in the calculation of vibrational entropies. Our aim is to
understand how the corresponding three physical properties affect vibrational spectra in general 
and vibrational entropy in particular.

\section{Results and discussion}
\label{Discussion}

It is our aim  to explore the correlation of vibrational entropy with secondary structure
content across the whole database. More precisely, the {\em fil rouge } of our analysis is to
check  whether entropic contributions might favor the formation of $\beta$-sheets at high
temperatures, thus in turn disfavoring $\alpha$-helices. In view of the positive correlation
existing between the $\alpha$ and $\beta$ contents,  (see again Fig.~\ref{f:alpha-beta}) we
shall consider in the following  the correlation of entropy with a global structural indicator
defined as ($\beta$-content $-$ $\alpha$-content), which we shall refer to as 
$\beta$-preference. Positive correlation of this quantity with entropy leads to a negative
correlation with free energy at high temperatures, that will in turn favor the formation of
$\beta$-like structures. Conversely, a negative correlation would  signal a preference
towards $\alpha$ helices. Importantly, in order to compare proteins of different lengths, we
shall always consider  {\em intensive} observables, namely entropies per residue.

A first analysis of the vibrational entropy of the PDBselect structures performed by
means of a standard ANM model yields somewhat unintuitive  results.
The intensive vibrational entropy $S_\mathrm{vib}/N$ 
does not correlate with the $\alpha$-helical content (correlation coefficient 0.14)
and  anti-correlates very weakly with the $\beta$-sheet  
content (correlation coefficient 0.34). This result, implying that $\beta$-sheet are mildly stiffer than
$\alpha$-helices, is manifestly at odds with the well documented rigidity of $\alpha$-helices.
We conclude that the ANM model, although reliable at the low-frequency end of the vibrational spectra
does not include a sufficient amount of detail  to realistically 
reproduce full vibrational entropies.

\subsection{The HNM model: quasi-additivity of vibrational spectra}

\begin{figure}[t!]
\centering
\includegraphics[clip,width=12 truecm]{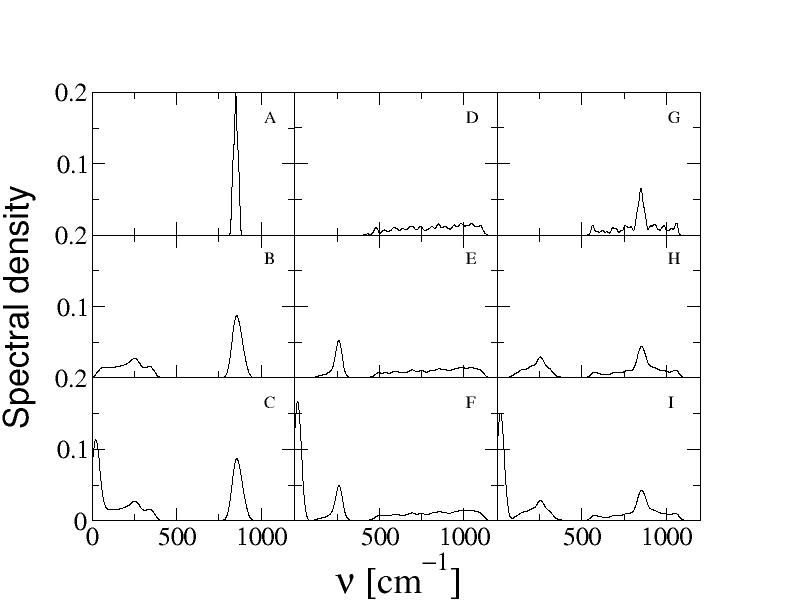}
\caption{HNM model without side chains and with equal masses. 
Spectral density after consecutive activations of
covalent bonds (upper row, $\beta=\gamma=0$), hydrogen bonds  (middle row, $\beta=0$) 
and van der Waals (bottom row) interactions for an $\alpha$-helical (panels A, B,
C $-$ PDB code \texttt{1KFM}), a $\beta$-barrel (panels D, E, F $-$ PDB code \texttt{1QJ9}) 
and a structure with equal $\alpha$ and $\beta$ content (panels G, H, I $-$ PDB code \texttt{1LSX}).
Parameters are: $R_{c}=6$ \AA, $\alpha=300$ kcal/mol/\AA$^2$, 
$\beta=30$ kcal/mol/\AA$^2$ and $\gamma=0.3$ kcal/mol/\AA$^2$.}
\label{f:densities}
\end{figure}

As a first trial towards increasing the level of detail, we consider the bare HNM model,
that is we introduce a three-level hierarchy of spring strengths in the framework 
of the standard ANM scheme as from point  (i) above.
Moreover, we shall switch on the different force constants one at a time first, in
order  to investigate the effects of the different physical interactions upon 
vibrational entropies. 
A first observation is that a roughly ten-fold separation in the intensity 
of force strengths is enough to induce quasi-additivity of the vibrational
spectrum. In Fig.~\ref{f:densities}, we show how the vibrational spectrum of sequences \texttt{1KFM}
(a purely $\alpha$-helical protein), \texttt{1QJ9} (a $\beta$-barrel) and \texttt{1LSX} 
(36\% $\alpha$ and 36\% $\beta$) 
change upon sequentially switching on the three different interactions. 

As a general remark, we observe that, for all types of structure, the bare covalent chain has
only $N$ modes with non-zero frequency.  
More in detail, we see that the structure  with dominant $\alpha$-helical character has a very distinctive 
peak at high frequency. Moreover, such  feature shows to be robust with respect to the nature of residue-residue
interactions. A peak is also seen to emerge in the $\beta$ structure as soon as the connectivity matrix
starts having its off-diagonal regions populated due to HBs and VW interactions.

\begin{figure}[t!]
\centering
\includegraphics[clip, width=12 truecm]{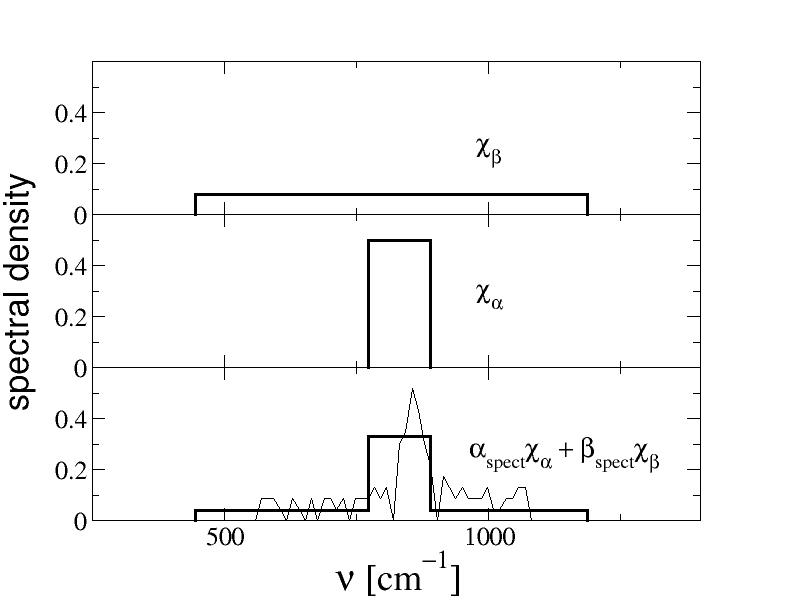}
\caption{HNM model without side chains and with equal masses. Estimating the  $\alpha$-content by fitting the high-frequency portion of the 
spectrum in a typical case (mixed $\alpha-\beta$ structure \texttt{1LSX}). 
The two representative functions $\chi_\alpha$ ($\alpha$-peak) and 
$\chi_\beta$ ($\beta$-band) are combined linearly  in the step function 
$\alpha_\mathrm{spect}\chi_\alpha+\beta_\mathrm{spect}\chi_\beta$, which is  least-square
fitted to the high-frequency portion of the spectrum. Parameters as in Fig.~\ref{f:densities}.}
\label{f:fitting}
\end{figure}

\begin{figure}[t!]
\centering
\includegraphics[clip, width=12 truecm]{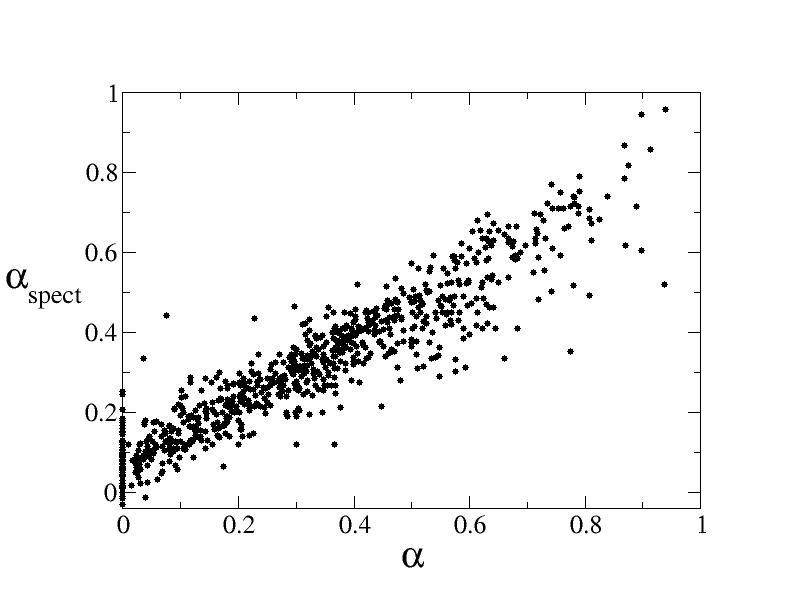}
\caption{HNM model without side chains and with equal masses. Correlation of the $\alpha$-content as  determined from the spectra, 
$\alpha_\mathrm{spect}$, and the actual $\alpha$-content as measured through the DSSP protocol 
for all structures from the PDBselect. Parameters as in Fig.~\ref{f:densities}.}
\label{f:formfactor-alpha}
\end{figure}

More quantitatively, we found that the high-frequency part of the spectra across 
all the PDBselect could be well reproduced by a linear combination of two normalized step functions 
$\chi_\alpha$ and $\chi_\beta$, with supports lying, respectively, in the $[750,1000]$ cm$^{-1}$ and 
$[500,750]$ cm$^{-1}$ intervals as shown in Fig.~\ref{f:fitting}. The two resulting coefficients
$\alpha_\mathrm{spect}$ and $\beta_\mathrm{spect}$
sum up to unity and might be regarded as a spectral estimate of the  actual $\alpha$ and
$\beta$ contents. At  a closer inspection,  it turns out that only $\alpha_\mathrm{spect}$ is a good 
approximation of the actual $\alpha$-content of the protein 
(see Fig.~\ref{f:formfactor-alpha}), while $\beta_\mathrm{spect}$ is observed to be large not only
when the protein has an high beta content but also when it has no pronounced secondary structural
features at all. In other words, while the $\alpha$-peak is a clear distinctive feature of
$\alpha$-helices, $\beta$-sheets are rather spectrally indistinguishable
from unstructured regions (see the appendix for a quantitative explication of the $\alpha$-peak).

\begin{figure}[t!]
\centering
\includegraphics[clip, width=12 truecm]{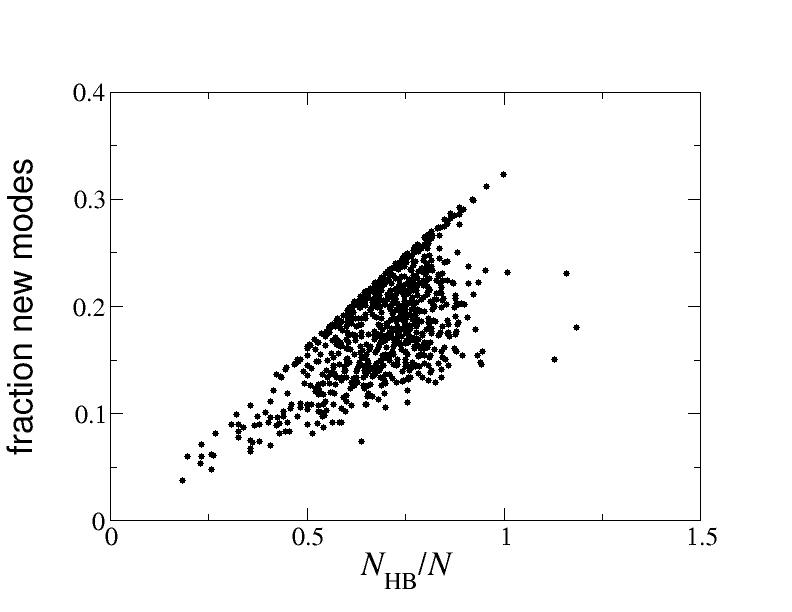}
\caption{HNM model without side chains and with equal masses. 
Correlation of the percentage of new modes contributed by adding HB interactions
with the number of hydrogen bonds per residue. Parameters as in Fig.~\ref{f:densities}.}
\label{f:percentage-newmodes}
\end{figure}

If we now turn the HBs on, vibrational spectra acquire a varying amount of new vibrational modes  
(between 5\% and 30\%), on average 18\% of the total number
of degrees of freedom on the entire PDB select. Perhaps not surprisingly, the quantity of such new 
modes turns out to be linearly correlated with the number of HBs (see Fig.~\ref{f:percentage-newmodes}).
Furthermore, we see that the appearance of the new HB part
of the vibrational spectrum only slightly modifies the spectral density at
higher frequencies (compare panels B and C, E and F  or H and I in Fig.~\ref{f:densities}).

Finally, adding the Lennard-Jones contribution, all the remaining modes are filled, with the
exception of the expected six zero-frequency modes corresponding to rigid translations and
rotations. Once again, due to the lower intensity of their driving interaction, the new LJ modes
occupy the low-frequency end of the spectrum and do not modify appreciably the spectral density at
higher frequencies.

\subsection{The full model: entropy- secondary structure correlations}

The marked differences in the spectra of $\alpha$ and $\beta$ proteins described so far through the 
HNM C$_\alpha$ model do not show up in the same clear-cut fashion if one examines how
the vibrational entropy depends on the type of secondary structure of the proteins: the intensive entropy 
shows no correlation at all with $\beta$-preference (data not shown). 

At a deeper thought, this is scarcely surprising, since the only detectable spectral differences between proteins 
of different secondary organization have already seen to be  
located at high frequencies, which in turn results in a  negligible contribution to vibrational entropy. 
What is more, that part of the spectrum is also the most sensitive to the structural
details of the system and consequently the least reliably captured by a coarse-grained
model. In order to test the solidity of our conjectures, we must then increase the level of 
descriptive accuracy of the model. 
We have devised two ways to accomplish this, by adding additional degrees of freedom representing 
the side chains or adding the full sequence of aminoacid masses.

\begin{figure}[t!]
\centering
\includegraphics[clip, width=12 truecm]{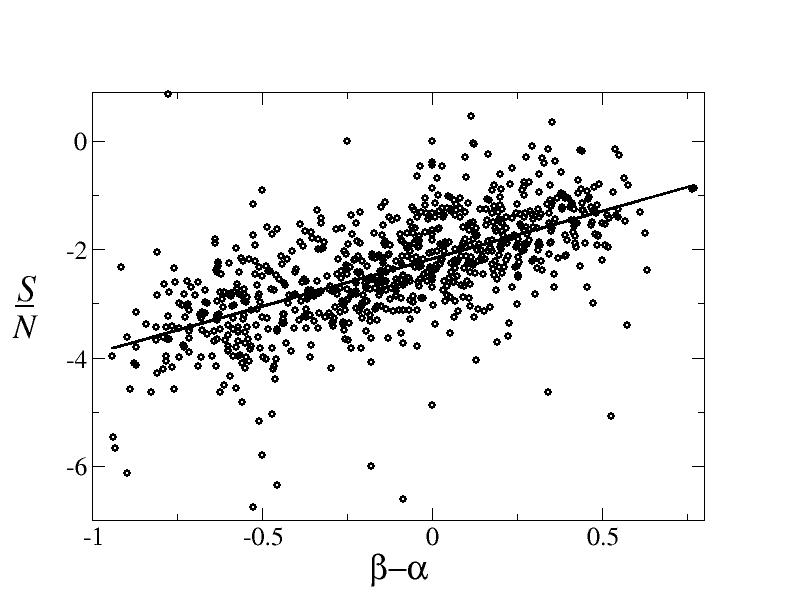}
\caption{Correlation of the total entropy per residue $S/N$
with $\beta$-preference in the weighted HNM model with side chains.
Parameters are: $R_{c}=6$ \AA, $\alpha=300$ kcal/mol/\AA$^2$, 
$\beta=30$ kcal/mol/\AA$^2$ and $\gamma=0.3$ kcal/mol/\AA$^2$.}
\label{f:correlationsSideChain}
\end{figure}

We found that both additional features independently produce the same effect, namely
destroy the clear $\alpha$-peak versus $\beta$-band picture illustrated above. However,
the spectral additivity is preserved and, all the more significantly, a good correlation between intensive 
vibrational entropy and $\beta$-preference appears, as can be clearly seen in 
Fig.~\ref{f:correlationsSideChain} where the two observables are scatter-plotted for the 
full, weighted HNM model with side chains.

A systematic analysis reveals that such correlation is shaped by a complex interplay between
the features that distinguish our model from the basic ANM scheme. In table~\ref{t:models}
the arousal of the correlation between entropy and $\beta$-preference is summarized by
reporting both the correlation coefficient and the slope of  the first order least-square fit
to the data. The two best correlations achieved are reported in bold. 

The best correlation is obtained when all levels of detail are present. In order to  quantify their
different contribution, it is useful to compute the average correlation coefficients  over the four
instances where a single feature is always present. For example, the average correlation coefficient
is only 0.27 when side-chain coordinates are included (last four rows of table~\ref{t:models}) and
0.29 in the four cases where true masses are taken into account (rows 2,4,6,8). On the contrary,
the average correlation coefficient of 0.39 obtained by sticking to realistic interaction strengths
points to a crucial role of the  protein dynamical heterogeneity in reproducing vibrational modes
associated  with different secondary structure motifs.

Overall, the same conclusions can be drawn by calculating the correlation drops resulting
from the individual elimination of single features from the complete model (last row in 
table~\ref{t:models}). In particular the importance of 
force hierarchy is confirmed (a correlation drop of about 0.5 from rows 8 to 6).
Moreover, this analysis reveals that considering equal masses also leads to the same 
correlation drop (rows 8 to 7). On the contrary, the elimination of side chains appears less 
traumatic with a limited loss of correlation (rows 8 to 4).  

In summary, concerning the interplay between secondary organization and vibrational properties,
the three levels of detail introduced can be ranked as follows: 
the hierarchy of physical interactions proves to be the most important feature, followed by 
mass heterogeneity and, finally, inclusion of side-chain (coarse-grained) degrees-of-freedom. 

\begin{table}[t!]
\caption{Summary of the effect of physical features of the HNM model on the 
correlation between vibrational entropy and $\beta$-preference: $\surd=$ feature on, 
$-$ feature off. The first row corresponds to the bare ANM model.}
\centering
\begin{tabular}{@{}ccc|rr}
\hline
 {\bf SC} & {\bf Hierarchy} & {\bf Mass} & {\bf Correlation}  & {\bf Slope} \\
    &           &      &  {\bf coefficient} &     \\
\hline\hline
&&&&\\
$-$  & $-$  & $-$  &      $-$0.239     & $-$0.546 \\
$-$  & $-$  & $\surd$ &      $-$0.123     &  $-$0.244 \\
$-$  & $\surd$ & $-$  & 	     0.261      & 0.588 \\
$-$  & $\surd$ & $\surd$ & {\bf  0.555}      & 0.954 \\
$\surd$ & $-$  & $-$  &    	0.253      & 0.639 \\
$\surd$ & $-$  & $\surd$ &    	0.106      & 0.322 \\
$\surd$ & $\surd$ & $-$  &    	0.106      & 0.514 \\
$\surd$ & $\surd$ & $\surd$ & {\bf  0.634}     & 1.747 \\
\hline
\end{tabular}
\label{t:models}
\end{table}

\begin{figure}[t!]
\centering
\includegraphics[clip, width=12 truecm]{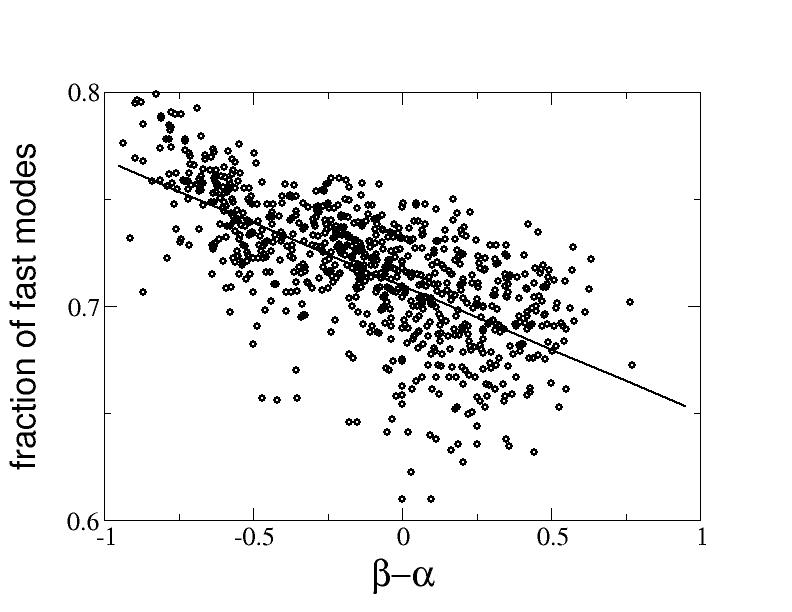}
\caption{Correlation between $f$, the fraction of modes determined by covalent and hydrogen bonds,
and $\beta$-preference over the PDBselect database using the full HNM model with side chains and 
proper mass sequences. 
The correlation coefficient is 0.44. Parameters are: $R_{c}=6$ \AA, $\alpha=300$ kcal/mol/\AA$^2$ and 
$\beta=30$ kcal/mol/\AA$^2$ and $\gamma=0.3$ kcal/mol/\AA$^2$.}
\label{f:fastmodes}
\end{figure}

This correlation reflects a tendency of $\beta$-rich architectures 
to host more low-frequency modes. This mechanism can better illustrated by switching off the 
weakest force constant in the complete model, that is shut down LJ bonds. 
In this case the $M'$ vibrational modes with non-zero frequency describe 
the oscillations of the network of interactions determined by HBs and covalent bonds alone, and 
represent all the oscillations with wavelength shorter than or equal to the typical size of
secondary structure motifs.  
The ratio $f=M'/M$, with $M$ the number of non-zero frequency vibrational modes in the full mode 
might be regarded as the fraction of modes determined by the HB and covalent interactions.
This assumption appears reasonable in view of the spectral quasi-additivity discussed above.
More precisely, we expect that zero-frequency modes will be filled when recovering LJ interactions, 
without significantly affecting the high-frequency portion of the vibrational spectrum. 
Fig.~\ref{f:fastmodes} shows that $f$ correlates negatively with the $\beta$-preference. 
Thus, protein structures with large  $\beta$-preference host fewer high-frequency modes and 
accordingly more low-frequency ones. This, in turn, leads to larger vibrational entropies for 
$\beta$-rich structures. 

More explicitly, we have shown that eliminating all contacts but those contributed by
secondary  structure motifs a clear separation of $\alpha$ and $\beta$-like vibrational spectra
emerges. This is further illustrated by switching off both Van der Waals interactions and HBs
(data not shown). In this case we no longer observe a correlation between $f$ and
$\beta$-preference. Thus, chain topology alone does not introduce any clear spectral signature 
of secondary structure organization.

\section{Conclusions and outlook}
\label{Conclusions and outlook}

In this work we have scrutinized a  wide database of non-homologous protein structures, with the 
aim of assessing to what extent vibrational 
entropy is a sensitive measure of organization at the secondary structure level. 
To this end, we have introduced a  minimally featured coarse-grained model, that we have 
coined Hierarchical Network Model (HNM). At variance with current, stat-of-the-art schemes, 
our  recipe  (i)  includes a hierarchy of physical interactions separating the strongest, short-range  forces
from the weaker, long-range ones, (ii) expands the number of degrees of freedom with the inclusion of
side-chain coordinates and (iii) accounts for the appropriate aminoacid  masses.

A thorough analysis of all the structures in the data set has allowed us to unveil and quantify
the correlation existing between the vibrational entropy of native folds and their specific 
secondary structure arrangement. 
More precisely, we found that {\em all} the three above-listed features of the HNM are essential in order to spotlight and 
quantify such correlation.  Remarkably, the statistical significance associated with the mixed spectral-structural signature
attains its maximum value only for the full-featured scheme. 
Of remarkable interest is the special role played by the requirement of force heterogeneity.
In fact, we have proved that the presence of a realistic hierarchy of force constants accounts for
the largest contribution to the observed correlation.

\bigskip 

As a final observation, it is interesting to discuss our results in the perspective of
the widespread, yet still much debated phenomenon of protein aggregation.
More precisely, concerning what physical effect could be held responsible for the overwhelming $\beta$ 
preference of mature peptidic aggregates~\cite{aggregation-review}.

The lack of clear signatures of the aggregation propensity at the 
sequence level~\cite{sequence-determinants} suggests that the phenomenon should be
governed by rather general laws, uniquely related to the dynamical properties of poly-peptide 
chains in solution. In particular, in view of the rather slow time-scales characteristic of 
aggregation, it is reasonable to postulate a mechanism of thermodynamic origin, 
that would favor the growth of structures rich in $\beta$-content under quite generic conditions. 
Of course, forces of enthalpic origin represent the strongest interactions along typical
aggregation pathways. However,  there is no reason a-priori for the strongest forces to also encode the observed
bias toward $\beta$-rich structures.
In fact, a composite structure developing from the aggregation of poly-peptides
would likely show a tendency toward non-specific organization at 
the secondary structure level, thus realizing  {\em minimal frustration} architectures~\cite{tubone}.

In view of the above facts, it is tempting to rationalize the aggregation pathway 
as {\em driven} by strong, non-specific forces but {\em biased} toward 
high content of $\beta$-type motifs by contributions to the total thermodynamic force that are weaker
in magnitude but favor $\beta$-rich architectures due to their higher density of vibrational modes at low frequencies.
Hence, based on the results reported in this paper,
one may speculate that the existing  bias toward  $\beta$-rich mature aggregation 
products  be provided by a free-energy gain in  vibrational entropy.
Following this speculation, the bulk of the free-energy changes 
occurring during aggregation would be mainly determined by increase of residue-residue 
contacts and decrease of solvent-exposed surface, 
while even tiny differences in vibrational entropy could be able to steer an aggregating
system toward a $\beta$-rich configuration.

\appendix\section{The $\alpha$-peak} 

In order to rationalize the emergence of the $\alpha$-peak in the HNM-C$_\alpha$
model, we diagonalize the Hessian matrix around a class of regular 
chain configurations that include both $\alpha$-helices and beta sheets: those 
characterized by fixed values of both the bond and dihedral angle.
The configurations of a linear polymer composed of $N$ consecutive
residues can be described by a vector
\begin{equation}
(\vec{x}_1,\ldots,\vec{x}_N)=(x_1,y_1,z_1,\ldots,x_N,y_N,z_N), 
\end{equation}
or ,alternatively,
once $\vec{x}_1$ is known, by means of the three
vectors $\vec{r}_{i}=\vec{x}_{i+1}-\vec{x}_{i}$ with $i=1,2,\ldots,N-1$.
Let us now define a configuration of known dihedral angle $\phi$ according to 
the following three rules,
\begin{equation}
\label{r}
r_i=|\vec{r}_{i}|=1,
\end{equation}
\begin{equation}
\label{bond}
\vec{r}_{i}.\vec{r}_{i+1}=\cos \theta 
\end{equation}
(bond angle must be constant) and
\begin{equation}
\label{dihedral}
\vec{r}_{i_1}\times\vec{r}_{i}\times\vec{r}_{i+1}=\sin^2 \theta \cos \phi
\end{equation}
(dihedral angle must be constant). Once $\vec{x}_{i_2}$, $\vec{x}_{i_1}$ and 
$\vec{x}_{i}$ are known, $\vec{x}_{i+1}$ can be found solving the system of two 
linear equations \ref{bond} and~\ref{dihedral} and than by imposing the constraint~\ref{r}.
In the simplest case, $N=4$, the shortest chain on which a dihedral angle
can be defined, one can easily show that the eigenvalues of the Hessian of the 
potential 
\begin{equation}
V=k(r_1-1)^2+k(r_2-1)^2+k(r_3-1)^2
\end{equation}
are nine null ones plus
\begin{equation}
(2-{\sqrt 2}\cos \theta) k, \, 2k, \, (2+{\sqrt 2}\cos \theta) k.
\end{equation}
In other words they only depend on the bond angle and they tend to peak 
at $2 k$ for $\theta=90^\circ$.
This last feature nicely explains the $\alpha$-peak and the $\beta$-band, being the 
average C$_\alpha$-C$_\alpha$ bond angle in an $\alpha$-helix just slightly lower than 
$90^\circ$, while it is around $50^\circ$ for a $\beta$-sheet.

\acknowledgements
This work has been partially supported by EU-FP6 contract 012835 (EMBIO).

%

\end{document}